\input{aipcheck}

\documentclass[
    ,final            
  ]
  {aipproc}

\layoutstyle{8x11double}

\begin{document}

\title{UVES-VLT High Resolution Spectroscopy of GRB080319B and GRB080330
Afterglows}


\author{Valerio D'Elia}{
  address={Inaf-Osservatorio Astronomico di Roma \& ASI Science Data Center}
}

\begin{abstract}
We study here the Gamma-Ray Burst (GRB) environment through the
analysis of the optical absorption features due to the gas surrounding
the GRB. In particular, we analyze high resolution spectroscopic
observations of GRB080319B and GRB080330 taken with UVES at the VLT,
starting 8m30s and 1.5hr after the GRB trigger, respectively. The
spectra show that the ISM of the GRB host galaxies are complex, with
several components contributing to the host absorption system. In
addition, we detect strong excited absorption lines, from which we
derive information on the gas distance from the site of the GRB
explosion. Under the assumption that the excited features are produced
by indirect UV pumping, we found that this distance results to be 2-6
kpc for GRB080319B and $280\pm50$ pc for GRB080330, meaning that the
power of the GRB radiation can influence the conditions of the
interstellar medium up to a distance of several hundred pc.

\end{abstract}

\maketitle


\section{Introduction}

For a few hours after their onset, Gamma Ray Bursts (GRBs) are the
brightest beacons in the far Universe, offering a superb opportunity
to investigate both GRB physics and high redshift galaxies.  Early
time spectroscopy of GRB afterglows can give us precious information
on the kinematics, geometry, ionization and metallicity of the
interstellar matter of GRB host galaxies up to a redshift z 4, and of
intervening absorbers along the line of sight.  High resolution
spectroscopy is important for many reasons: (i) absorption lines can
be separated into several components belonging to the same system;
(ii) the metal column densities can be measured through a fit to the
line profile for each component; (iii) fine structure and other
excited lines can be resolved.

\begin{figure}
\centering
\includegraphics[angle=-90,width=9cm]{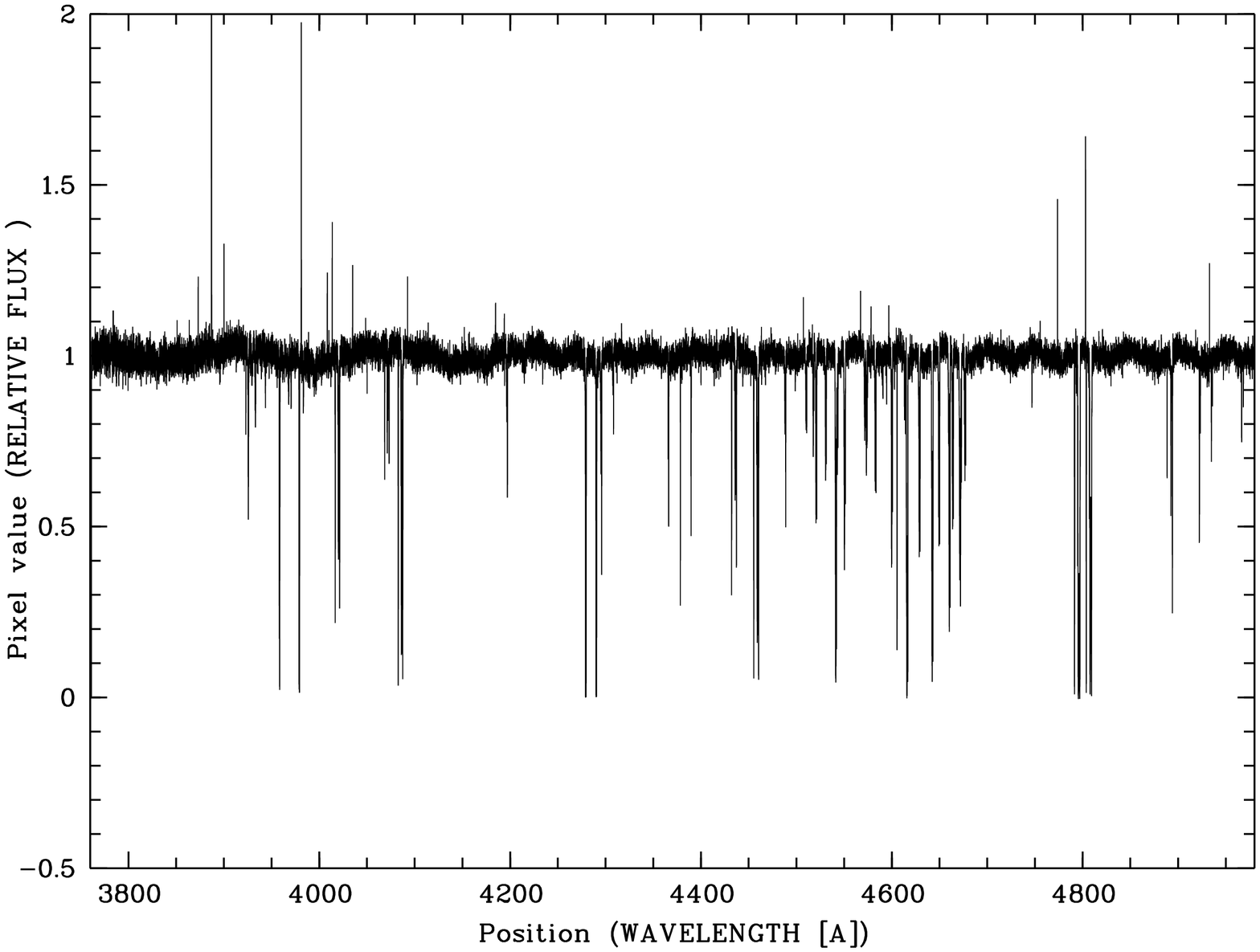}
\includegraphics[angle=-90,width=9cm]{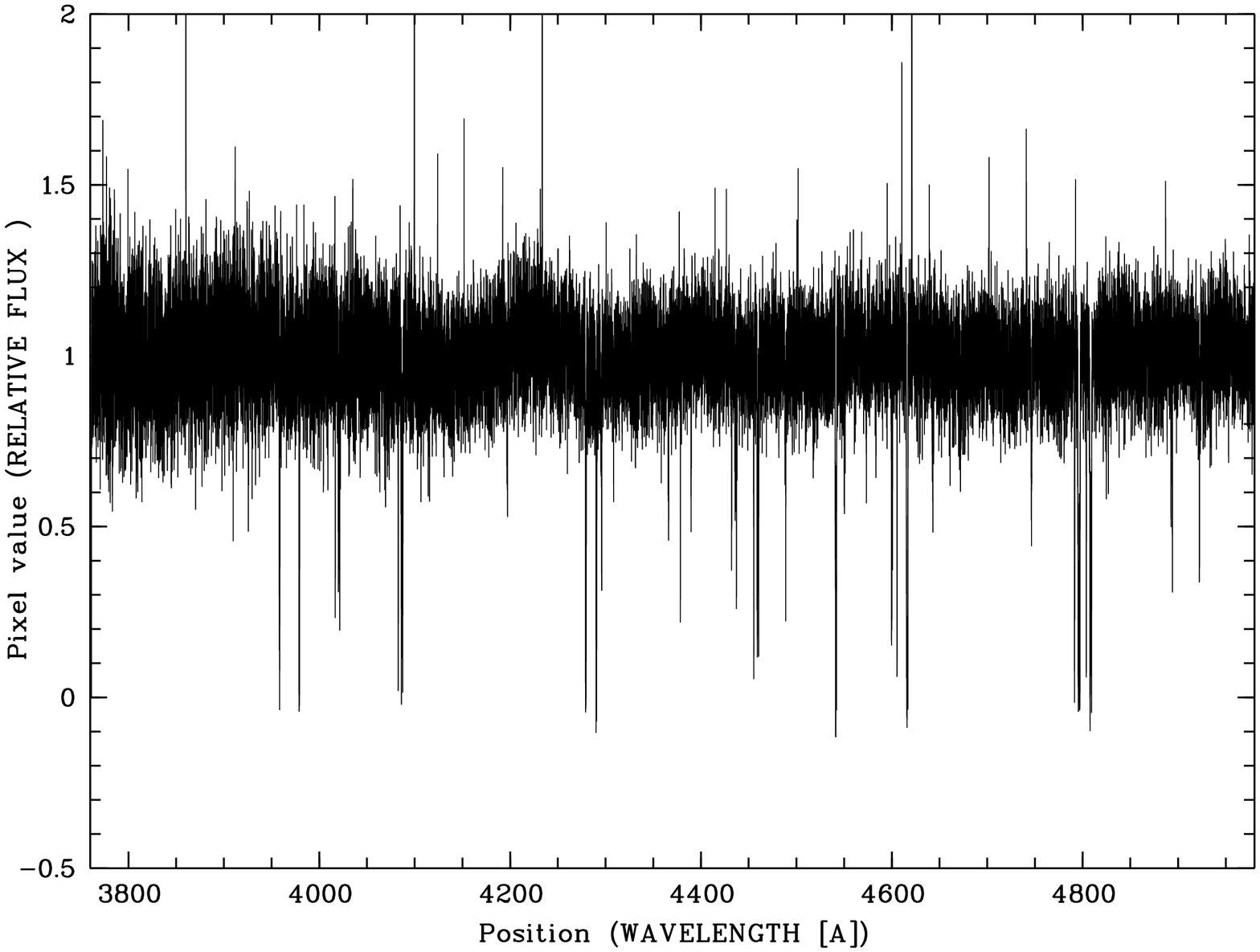}
\caption{The high resolution spectrum of GRB080319B taken 
with UVES 8,5m (left) and 1.9hr (right) after the trigger.}
\end{figure}

\begin{figure}
\centering
\includegraphics[angle=-90,width=7.5cm]{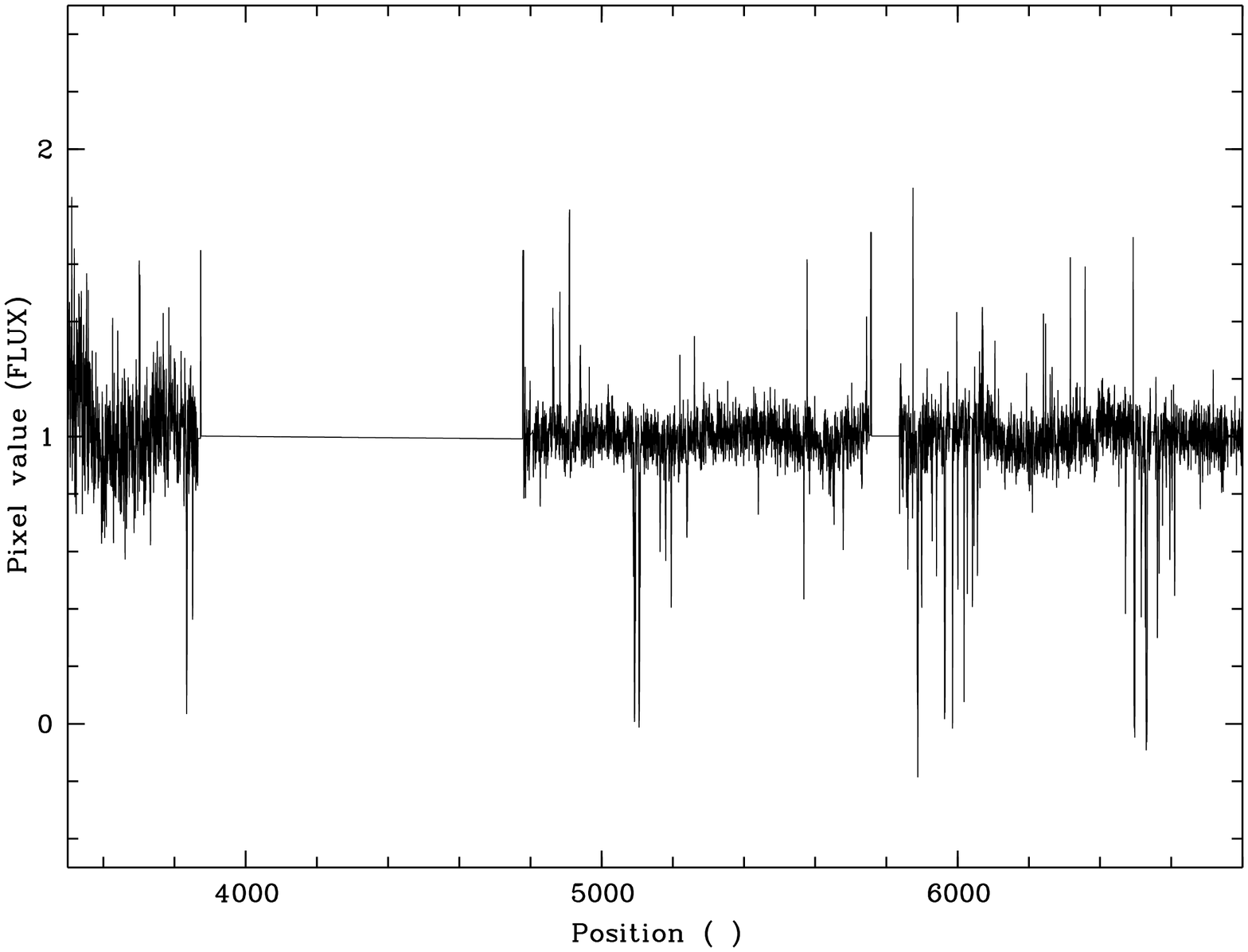}
\caption{The high resolution spectrum of GRB080319B taken 
with UVES 8,5m (left) and 1.9hr (right) after the trigger.}
\end{figure}

\section{Observations}
GRB080319B. This is the brightest GRB ever, visible
worldwide and naked-eye in the first seconds. The afterglow
has been observed from the very beginning of
the event. UVES was on target 8m30s later, when the
magnitude was R=12. This observation was followed by
two more exposures 2 and 3 hours later (Fig. 1).We find
five systems at z=0.937 (GRB host), 0.76, 0.71, 0.57 and
0.53.

GRB080330 afterglow was observed with UVES 1.6
hours after the trigger (Fig. 2). We find three main absorption
systems at z=1.51 (GRB host), 1.02 and 0.82
The resolution of the spectra reaches 7.5 km/s in the
observer frame. The S/N is 3-6 for GRB080330 and up
to 50 in the first observation of GRB080319B.

\begin{figure}
\centering
\includegraphics[angle=-90,width=9cm]{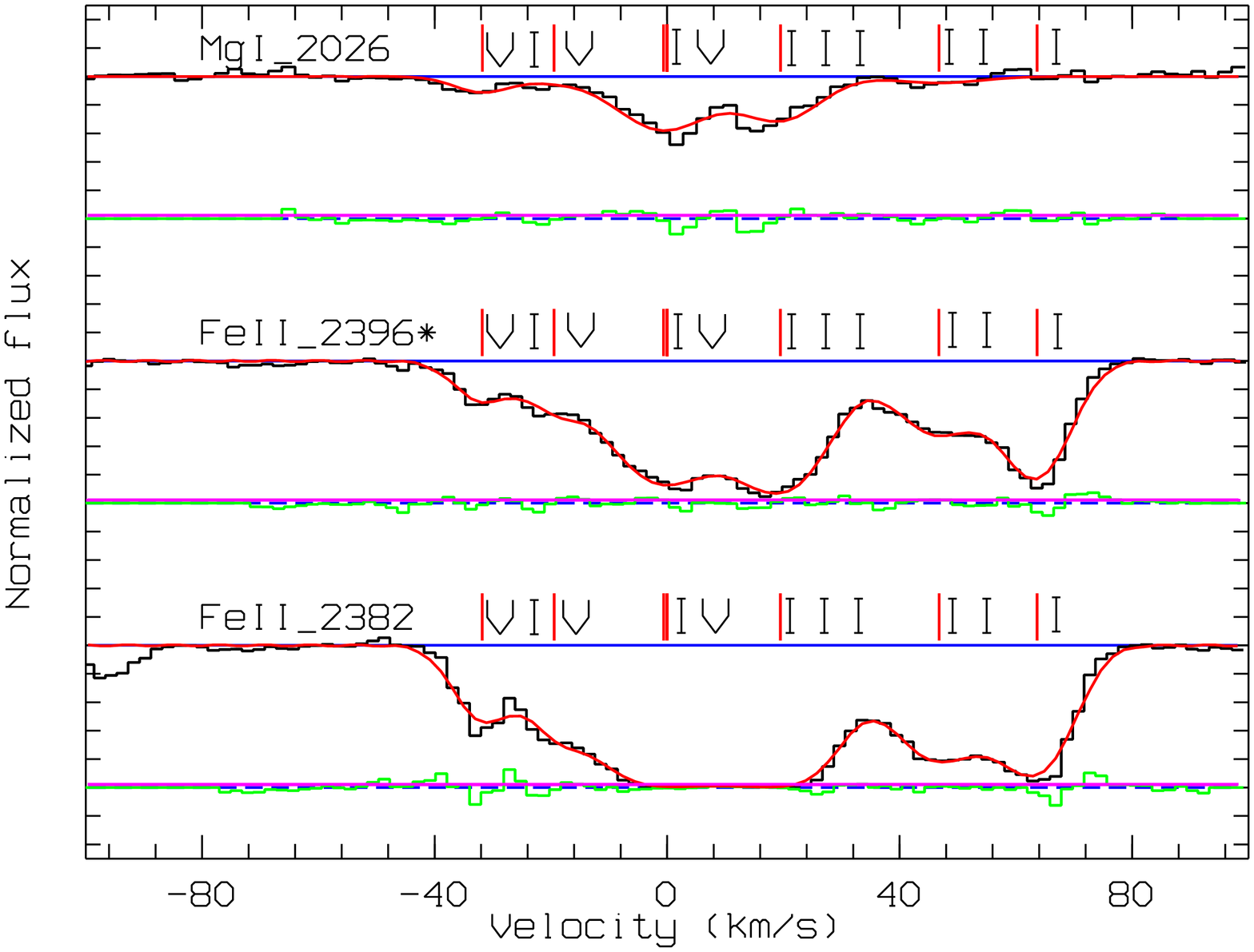}
\includegraphics[angle=-90,width=9cm]{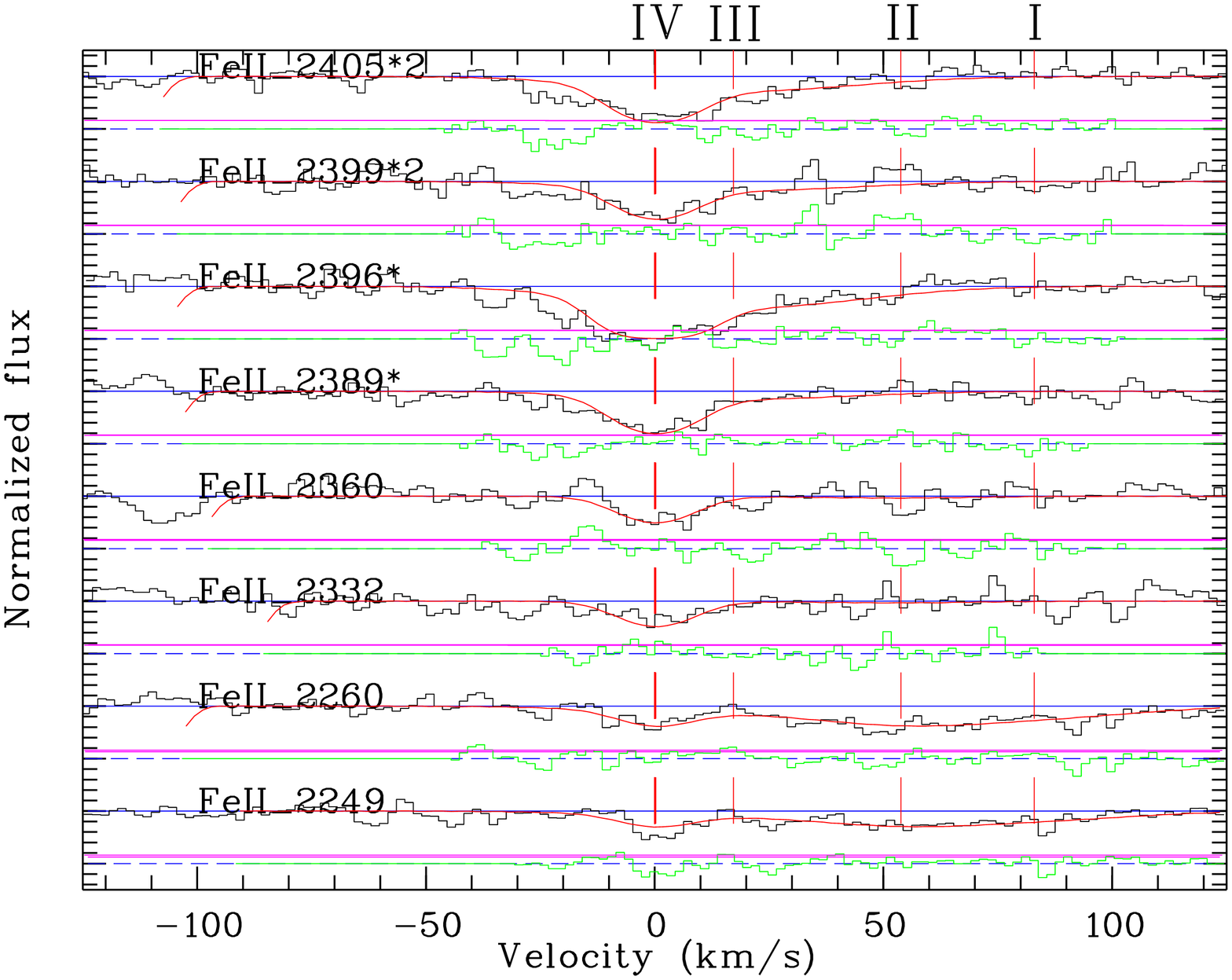}
\caption{Left panel: the FeII 2382, FeII 2396* (fine structure) and MgI
2026 absorption lines in GRB080319B. Right panel: several ground
(2249, 2260), fine structure (2389, 2396, 2399, 2405) and excited
(2332, 2360) FeII lines in GRB080330. Six and four absorption
components have been identified in the two GRB circumburst
environment, respectively.}
\end{figure}

\begin{figure}
\centering
\includegraphics[angle=-90,width=9cm]{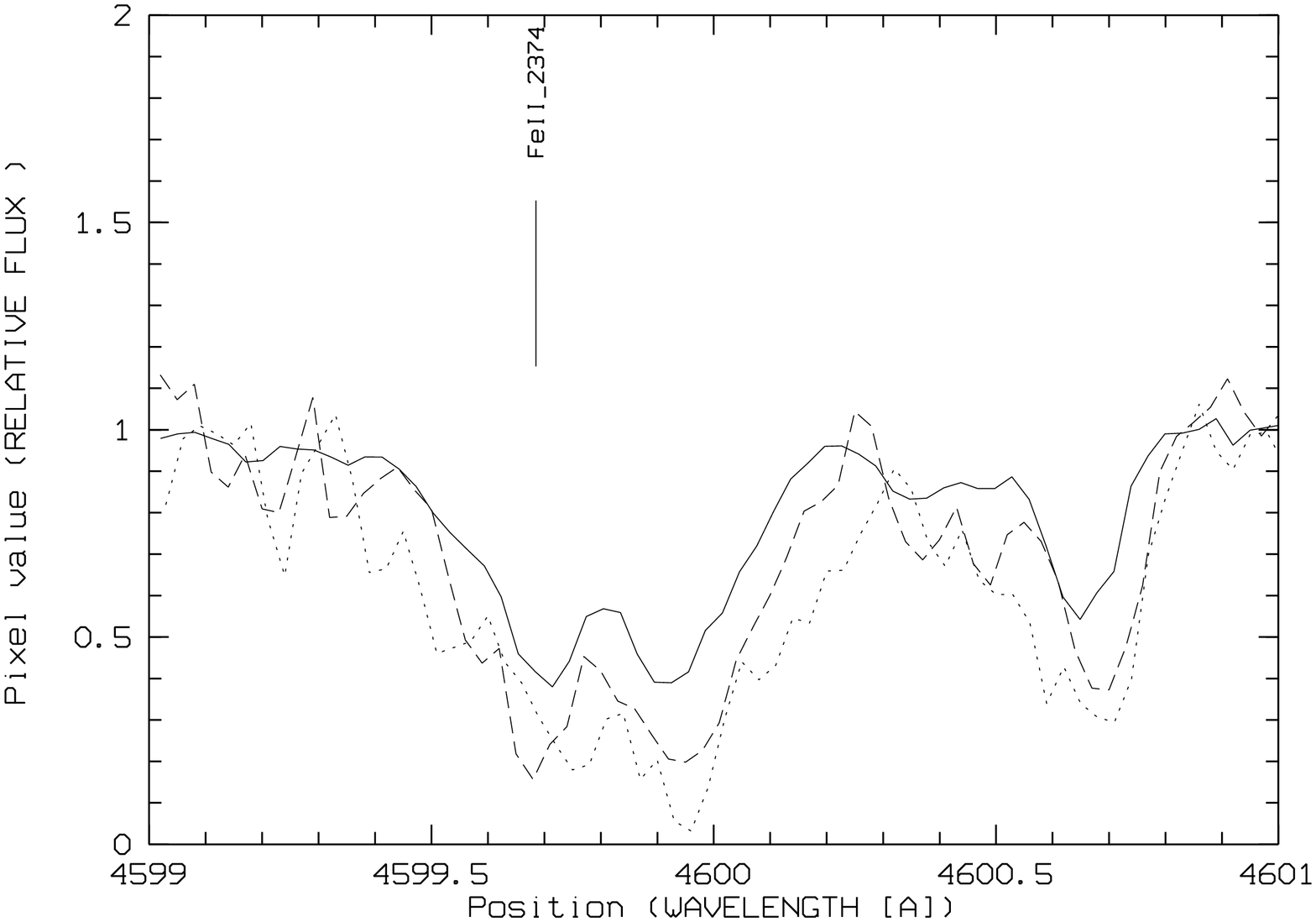}
\includegraphics[angle=-90,width=9cm]{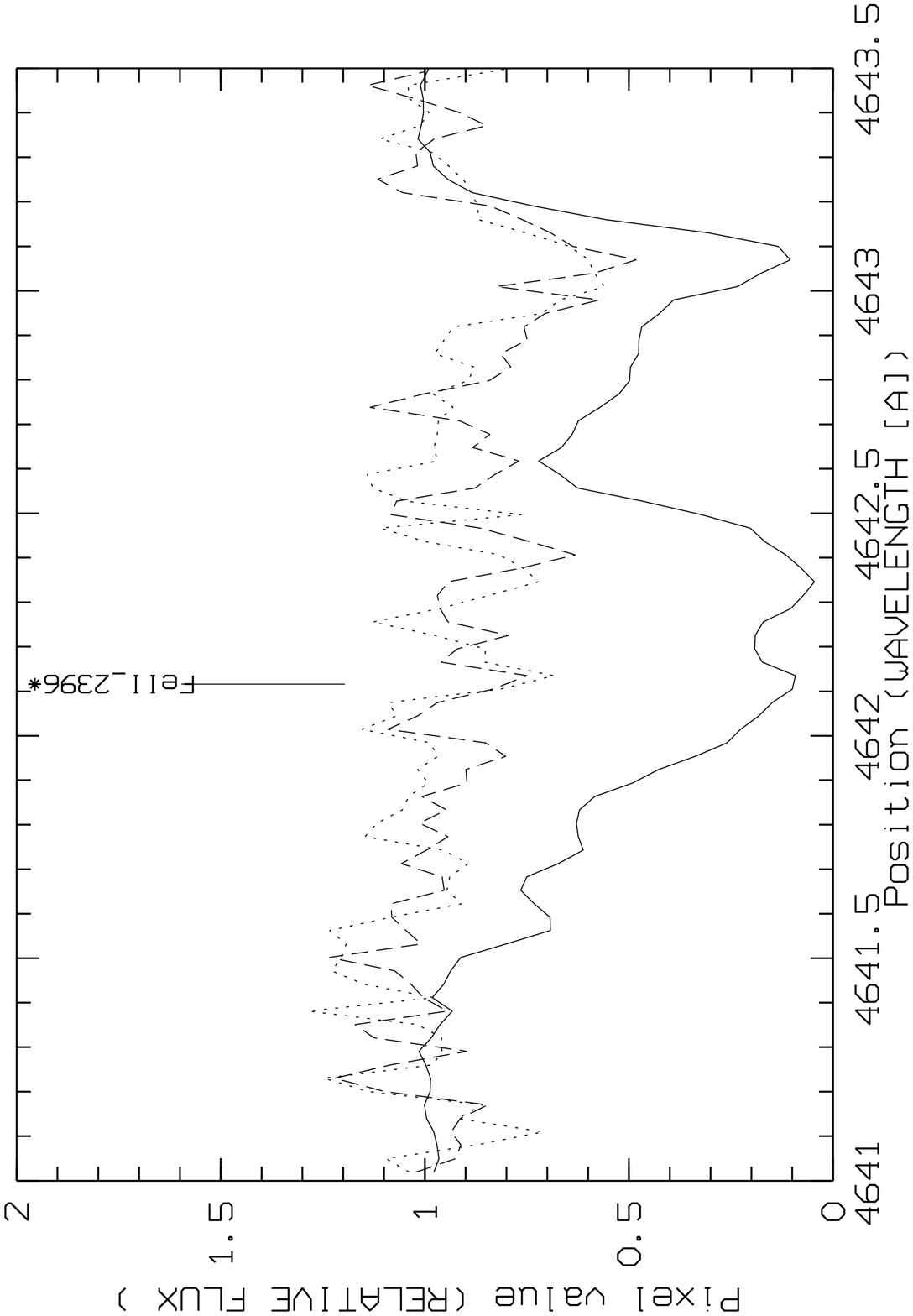}
\caption{ The FeII 2396 fine structure (left) and Fe II 2382 ground
state (right) lines for GRB080319B. Solid, dashed and dotted lines
represent the first, second and third UVES observation,
respectively. Note the strong variation of the fine structure line
while the ground state one is almost constant.}
\end{figure}

\begin{figure}
\centering
\includegraphics[angle=-0,width=9cm]{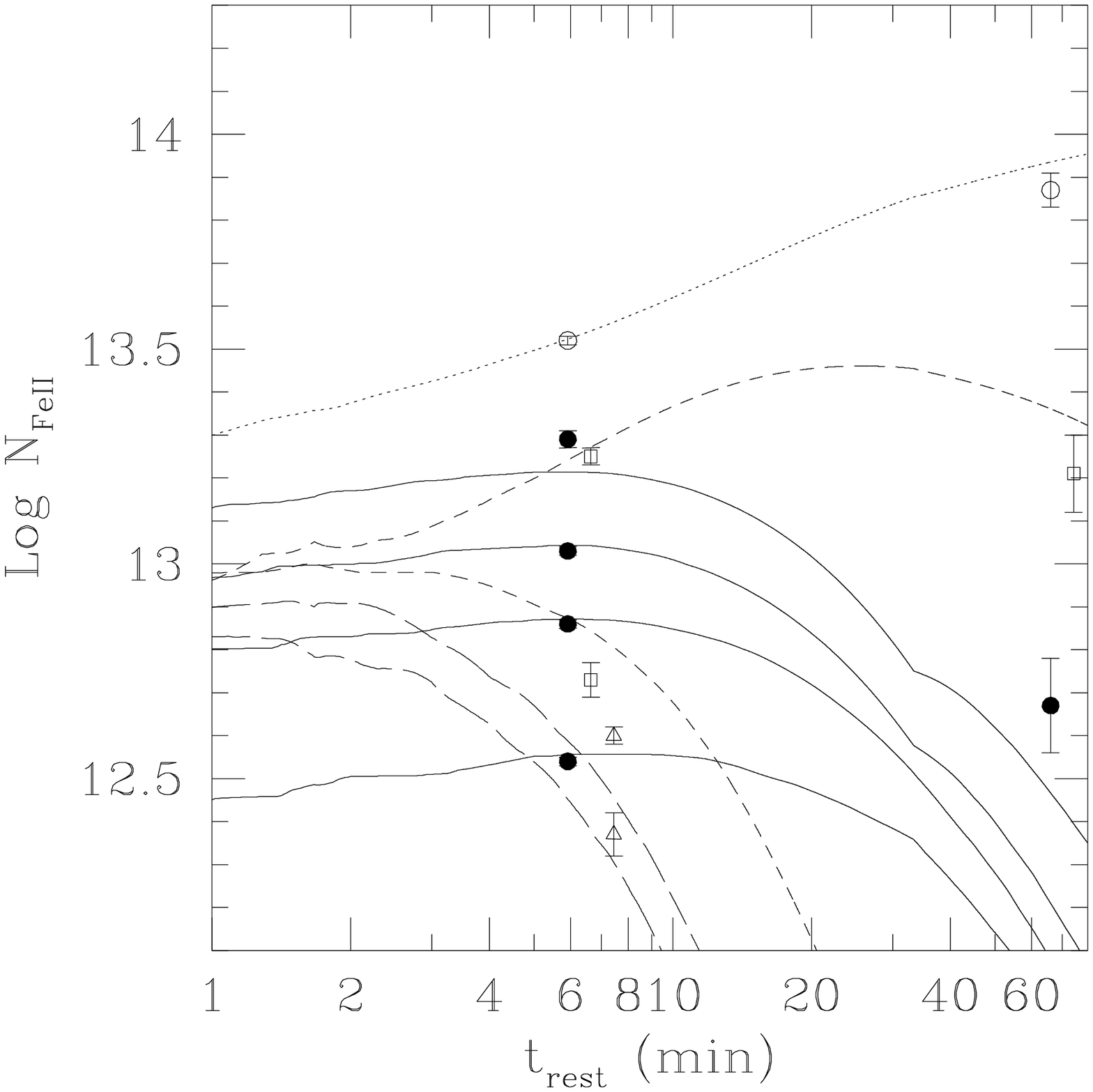}
\includegraphics[angle=-0,width=9cm]{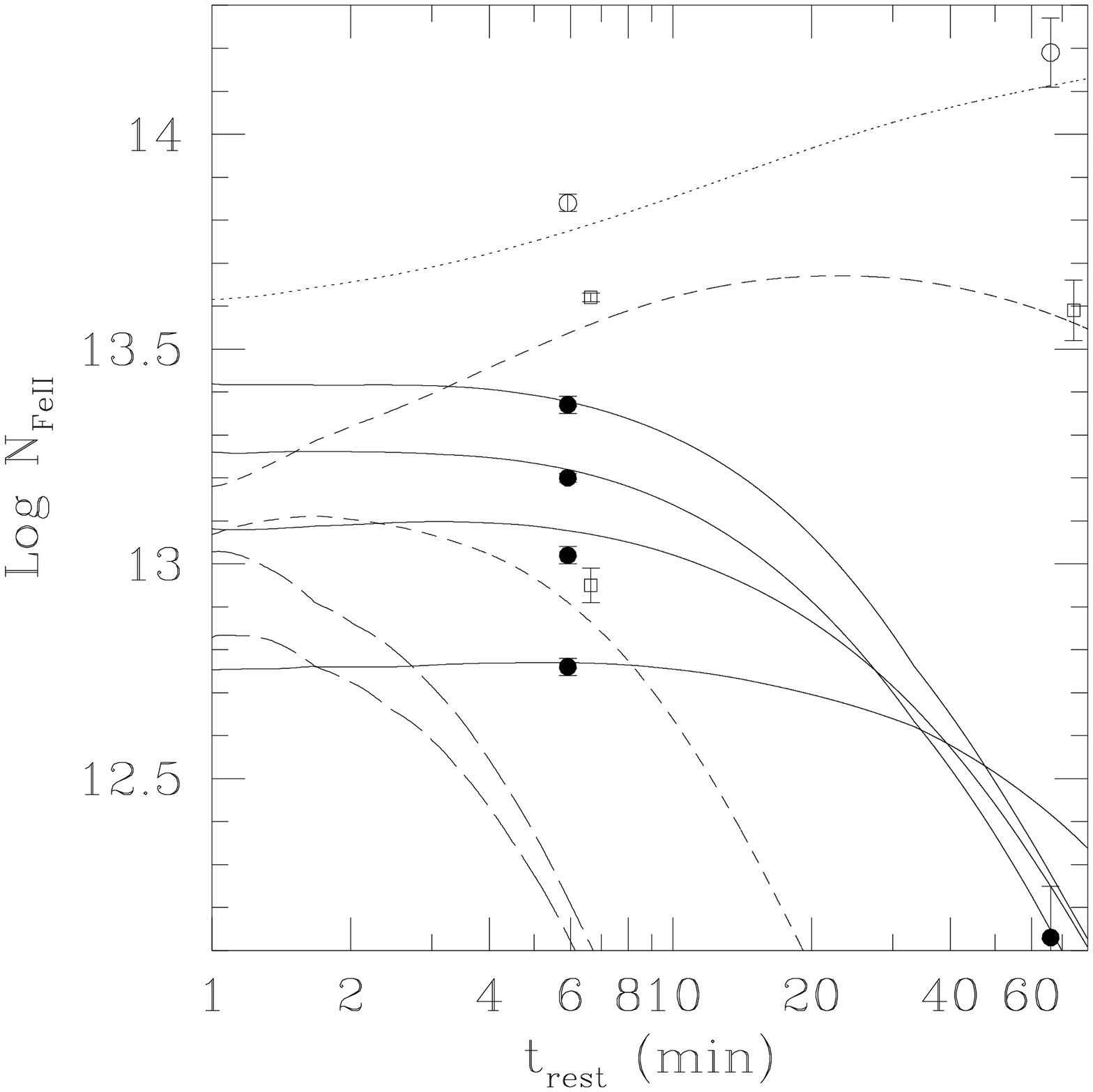}
\caption{Comparison between the observed column densities in the FeII
levels of GRB080319B and that predicted by our photo-excitation
code. Left panel refers to component I, for which a distance of 2 kpc
from the GRB is predicted. Right panel is for component III and a
distance of 6 kpc is found.}
\end{figure}

\begin{figure}
\centering
\includegraphics[angle=-0,width=7.5cm]{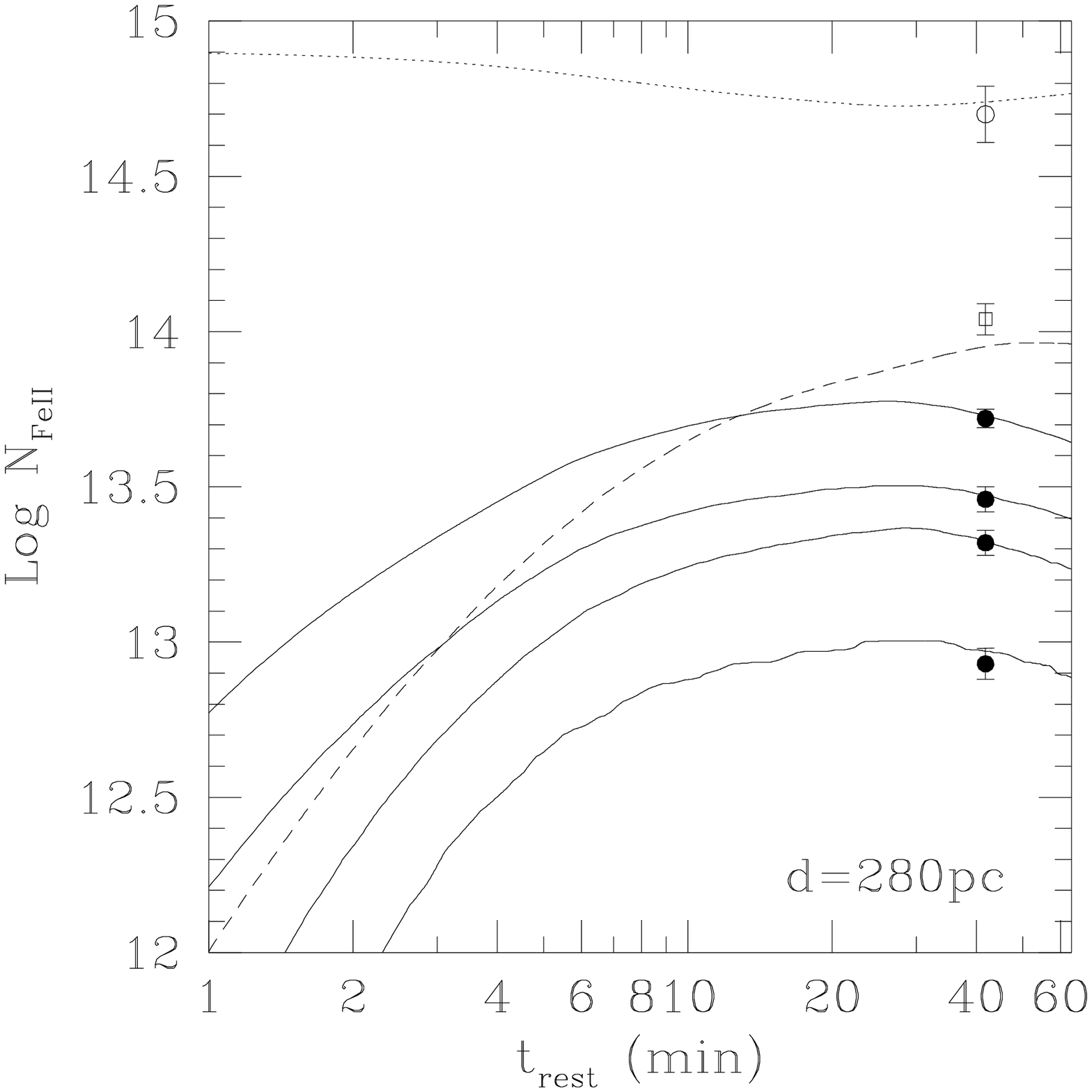}
\caption{Comparison between the observed column densities in the FeII
levels of GRB080330 and that predicted by our photo-excitation
code for component IV. A distance of $280 \pm 50$ pc
from the GRB is predicted.}
\end{figure}

\section{Analysis}

The features were analysed using the line fitting program FITLYMAN,
part of theMIDAS data reduction software package. FITLYMAN allows for
the simultaneous fitting of multiple absorption/emission systems. For
each absorption system several lines spread over the entire spectral
range covered by the UVES observations were fitted together. The
strongest absorption lines are observed in the host galaxy systems,
which are the only ones considered in the following. Six and four
components are necessary in order to obtain a satisfactory fit for the
line profiles of the absorber at the redshift of GRB080319B and
GRB080330, respectively (Fig. 3). Thus, as observed in previous high
resolution spectroscopy of GRBs (Fiore et al. 2005, D'Elia et
al. 2007, Piranomonte et al. 2008), the circumburst absorber shows
evidence of a clumpy structure, consisting of multiple shells.

\section{Excited Lines and their variability}

Fine structure lines and other excited features, belonging to the
FeII, NiII and SiII levels have been identified in these GRBs. Excited
lines can be produced both by collisional or radiative
processes. Vreeswijk et al (2007), using multi-epoch high resolution
spectroscopy, observed variability in the fine structure FeII lines of
GRB060418, which is a clear signature of indirect UV pumping exciting
such features. In other words, the decreasing UV flux coming from the
GRB excites the higher FeII levels with lesser and lesser
efficiency. An even stronger variability has been observed in
GRB080319B, where a variation of a factor of 4-20 in the FeII excited
lines have been detected in less than 1 hour rest frame, while the
ground state lines remained almost constant (Fig. 3).

\section{Gas distance form the GRBs}

Due to this strong variability in the excited features observed in
GRB080319B, we can safely assume that UV pumping is the responsible
for the production of these lines, and we can estimate the distance of
the gas from the GRB explosion site. First of all, we can state that
component 1 is the closest to the GRB, since it maintains a
significant absorption from the excited levels even at later times
(Figs. 4 and 5). A more quantitative approach needs the comparison of
the observed data with the results from a time dependent
photo-excitation code. We built a code that computes the column
densities of more than a hundred FeII levels as a function of an
incoming UV flux decreasing with time. Once the lightcurve of
GRB080319B has been used as input for this code, we can estimate the
distance of the absorbers from GRB080319B (Fig. 5).We find that the
gas of component I is 2 kpc away from the source, while that of
component III is at 6 kpc. This is a surprising result, since it
reveals a possible extragalactic origin of the gas clouds at the host
galaxy redshift, and tells us that the powerful GRB/afterglow emission
can fully ionize the intervening gas up to the kpc scale.  Even if we
do not have multi-epoch data for GRB080330, we can assume that UV
pumping is at work also for this GRB in the production of the excited
lines. Under this hypotesis, component IV results the closest to the
GRB, since it is the only one featuring excited levels
(Fig. 3). Moreover, we can compare again the results from the
photoexcitation code with the observed column densities (Fig. 6). We
find that for GRB080330 the distance of the closest absorber is
$280 \pm50$ pc, a smaller distance than in GRB080319B, possibly due to
the weaker flux of this GRB.

\section{Conclusions}

The absorption spectra of GRB 080319B and 080330 confirm that the GRB
afterglows are extremely complex, featuring several systems at
different redshifts. The host galaxy systems are clumpy, and are
constituted by 6 and 4 components, respectively. Such components can
be separated only with high resolution spectroscopy, which is thus the
only effective tool to allow for a detailed study of the GRB
surrounding medium. Among the host galaxy absorption features, fine
structure and other excited lines play a fundamental role. In
particular, the strong variability observed for these features in
GRB080319B allows to conclude that their production mechanism is UV
pumping. Moreover, the distance of the absorber from the GRB can be
estimated by comparison with time-dependent photo-excitation codes,
and the result is 2-6 kpc, i.e., the absorber can possibly have an
extragalactic origin. Assuming also for GRB080330 that UV pumping is
at work, we derive a distance of $280 \pm 50$ pc of the absorbing gas
from this GRB. This is a considerably smaller distance, possibly due
to the lower flux level of GRB080330. Nevertheless, it confirms that
GRBs influence their environment up to distances of several pc.

More details on this work can be found in D'Elia et al (2009 a\&b)


\end{document}